\documentclass[preprint]{elsarticle}
\usepackage[latin1]{inputenc}
\usepackage{color}
\usepackage[T1]{fontenc}
\usepackage[ english]{babel}
\usepackage[dvips]{graphicx}
\usepackage{amsmath}
\usepackage{palatino}
\usepackage{fancyhdr}
\begin{document}

\pagenumbering{arabic}

\begin{frontmatter}

\title{Using quantum key distribution for cryptographic purposes: a survey\tnoteref{t1}}
\tnotetext[t1]{This document is the fruit of a collaborative effort initiated within the FP6 Trust and Security European integrated project SECOQC (IST-2002-506813). It is based for a large part on the SECOQC Crypto White Paper \cite{cryptowp} that had been released in 2007.}

\author[tpt,sqn]{R.All\'eaume\corref{cor1}}
\author[gap]{C. Branciard}
\author[umas]{J. Bouda}
\author[trt]{T. Debuisschert}
\author[usurr]{M. Dianati}
\author[gap]{N. Gisin}
\author[ubris]{M. Godfrey}
\author[cnrs]{P. Grangier}
\author[arc]{T. L\"anger}
\author[iqc]{N. L\"utkenhaus}
\author[arc]{C. Monyk}
\author[tc]{P. Painchault}
\author[arc]{M. Peev}
\author[arc]{A. Poppe}
\author[crypt]{T. Pornin}
\author[ubris]{J. Rarity}
\author[ethz]{R. Renner}
\author[idq]{G. Ribordy}
\author[tpt]{M. Riguidel}
\author[umont]{L. Salvail}
\author[tosh]{A. Shields}
\author[lmu]{H. Weinfurter}
\author[uvie]{A. Zeilinger}

\cortext[cor1]{Editing author: {\tt romain.alleaume@telecom-paristech.fr}}

\address[tpt]{Telecom ParisTech $\&$ CNRS LTCI, Paris, France}
\address[sqn]{SeQureNet SARL, Paris, France}
\address[gap]{University of Geneva, Switzerland}
\address[umas]{Masaryk University, Brno, Czech Republic}
\address[trt]{Thales Research and Technology, Orsay, France}
\address[usurr]{University of Surrey, Guildford, UK}
\address[ubris]{ University of Bristol, United Kingdom}
\address[cnrs]{CNRS, Institut d'Optique, Palaiseau, France}
\address[arc]{Austrian Research Center, Vienna, Austria}
\address[iqc]{Institute for Quantum Computing, Waterloo, Canada}
\address[tc]{Thales Communications, Colombes, France}
\address[crypt]{Cryptolog International, Paris, France}
\address[ethz]{Eidgen\"ossische Technische Hochschule Z\"urich, Switzerland}
\address[idq]{Id Quantique SA, Geneva, Switzerland}
\address[umont]{Universit\'e de Montr\'eal, Canada}
\address[tosh]{Toshiba Research Europe Ltd, Cambridge, United Kingdom}
\address[lmu]{Ludwig-Maximilians-University  Munich, Germany}
\address[uvie]{University of Vienna, Austria}


\begin{abstract}
 The appealing feature of quantum key distribution (QKD), from a cryptographic viewpoint, is the ability to prove the information-theoretic security (ITS) of the established keys. 
As a key establishment primitive, QKD however does not provide a standalone security service in its own: the secret keys established by QKD are in general then used by a subsequent cryptographic applications for which the requirements, the context of use and the security properties can vary.
It is therefore important, in the perspective of integrating QKD in security infrastructures, to analyze how QKD can be combined with other cryptographic primitives.
The purpose of this survey article, which is mostly centered on European research results, is to contribute to such an analysis. We first review and compare the properties of the existing key establishment techniques, QKD being one of them. We then study more specifically two generic scenarios related to the practical use of QKD in cryptographic infrastructures: 1) using QKD as a key renewal technique for a symmetric cipher over a point-to-point link ; 2) using QKD in a network containing many users with the objective of offering any-to-any key establishment service. We discuss the constraints as well as the potential interest of using QKD in these contexts.  We finally give an overview of challenges relative to the development of QKD technology that also constitute potential avenues for cryptographic research. 
\clearpage
\end{abstract}


\end{frontmatter}

\tableofcontents

\section{Introduction}

In recent years quantum cryptography has been the subject of
strong activity and rapid progress \cite{GisinRevModPhys,
LutkQC, ScaraniRMP}, and it is now extending its activity to pre-competitive
research \cite{Secoqc} and to commercial products \cite{ qkdstartups}.
Nevertheless, the fact that quantum key distribution (QKD) can play a useful role in practical cryptography is sometimes considered with
skepticism \cite{WhyQC, Sh03, Sh07, Sh08} and cannot therefore been taken for granted. Analysing the practical
cryptographic implications of QKD is indeed a
complex task that requires a combination of knowledge that usually
belongs to separate academic communities, ranging from classical
cryptography to the foundations of quantum mechanics and network
security. Little work has so far been published on this issue, although \cite{VABook} may be considered as a pioneering contribution on that matter. This review article tries to identify in which contexts QKD can be useful, in addition to the scientifically well-established classical cryptographic primitives. 

The logical construction in the next three sections of this paper paper is to analyze the use of QKD, as a cryptographic primitive, for
different purposes, reflecting the first three layers of the OSI network model.
\begin{enumerate}
\item Secret key agreement (performed in the case of QKD at the physical layer). 
\item Secure payload transmission built on top of a key agreement scheme (secure link layer cryptographic primitive).
\item Secret key agreement over a global network composed of multiple users (network layer cryptographic primitive).
\end{enumerate}
 The paper is thus organized as follows:
In Section \ref{sec:KD}, we provide a survey of secret key agreement techniques and discuss some of their strengths, weaknesses, and
relative advantages.  In Section \ref{sec:MP}, we discuss the security and the performance of different secure payload transmission primitives that can be built on top of QKD, and that can be used to secure point-to-point communication links.
 In Section \ref{sec:networks}, we consider the use of QKD in a network context. We discuss previous works on QKD networks and also describe the cryptographic operation of such networks and in particular their initialization, that requires the distribution of pre-shared secrets. Finally, in  Section \ref{sec:futureD} we widen the scope of this survey paper by discussing some future research directions that could benefit from active collaboration between the quantum and the classical cryptography communities: the study of side-channels and of material security, the study of post-quantum-computing cryptography, the use of QKD networks as a strong building block for new network security protocols and the development of unified cryptographic standards and evaluation methods for quantum and classical cryptography.

\section{Secret key agreement} \label{sec:KD}

Cryptography has for a long time conformed to the idea that the
techniques used to protect sensitive data  had themselves to be
kept secret. Such principle, known as ``cryptography by
obscurity'' has however become inadequate in our modern era.
The cryptography that has developed as a science in the 1970s and
1980s  \cite{EcryptChallenges} has allowed us to move away from this
historical picture and most of the modern cryptographic systems
are now based on publicly announced algorithms while their
security lies in the use of secret keys.

 Distributing keys among a set of legitimate users while
guaranteeing the secrecy of these keys with respect to any
potential opponent is thus a central issue in cryptography, known as
the {\it secret key agreement problem}.

 There are currently five families of cryptographic methods that can be used to solve the secret key agreement problem between distant users:
\begin{enumerate}
\item Classical ITS schemes 
\item Classical computationally secure public-key cryptography
\item Classical computationally secure symmetric-key cryptographic schemes
\item Quantum key distribution
\item Trusted couriers
\end{enumerate}
We will present  how each of these cryptographic families
can provide solutions to the key agreement problem and
discuss, in each case, the type of security that can be provided.
We will also consider a sixth type of secret key agreement schemes:
hybrid schemes built by combining some of the methods listed
above.

\subsection{Classical information-theoretically secure key agreement schemes} \label{subsec:CITKE}

A cryptosystem is information-theoretically secure (ITS) if its
security derives purely from information theory. That is, it makes
no unproven assumptions on the hardness of some mathematical
problems, and is hence secure even when the adversary has
unbounded computing power. The expression ``unconditional
security'' is a synonym of ``information-theoretical security''
and is more widely used in the cryptographic literature. 

Studying the question of classical ITS secret key agreement (CITSKA) requires us to go back to the foundations of information-theoretic security, which builds on Shannon's notion of perfect secrecy \cite{Shannon}. In seminal papers, Wyner \cite{Wyner} and later Csisz\`ar and K\"orner \cite{CK} proved that there exist channel codes guaranteeing both robustness to transmission errors and an arbitrarily small degree of information leakage towards non-authorized parties eavesdropping on the communications performed on the channel.  CITSKA is possible in the wire-tap configuration, as long as the legitimate users have access to a common source of randomness through classical channels that are less noisy than the channel the eavesdropper has access to \cite{CK}. The results obtained by Csisz\`ar and K\"orner generalize the framework in which CITSKA is possible: they show that whenever  two parties have in their possession correlated strings of classical data that exhibit more correlation between them than with any string that could be in the possession of an eavesdropper, then ITS secret key agreement is possible. As we shall see in \ref{subsec:QKD}, the use of a quantum channel  and of an appropriate protocol is a practical solution in order to obtain such correlated strings of classical data.

There are however also secret key agreement schemes that can exploit the ideas developed in \cite{CK} and that can be implemented
within the framework of classical information theory. Such CITSKA schemes however need to rely on some specific extra assumptions,
limiting the power of the eavesdropper in order to be ITS.  Christian Cachin and Ueli Maurer  \cite{Cachin97} 
demonstrated that  CITSKA is possible in the bounded-storage model, in which the adversaries can only store a limited amount of data. Introducing the idea of advantage distillation, Maurer later generalized the previous models and showed that CITSKA is possible over a wide class of classical channels \cite{Maurer93}.

 \subsection{Classical public-key cryptography and secret key agreement}  \label{subsec:asym}

Public-key cryptography foundations rest on the difficulty of
solving some mathematical problems for which no efficient
algorithms are known. The computing resources needed to solve
these problems  become totally unachievable when long enough keys
are used. Public-key cryptographic systems thus rely on what is
called ``provable computational security''. Public-key cryptography is however
not unconditionally secure: there is no proof that the problems on which it is based are intractable or even that that their complexity is not polynomial.

 Public-key algorithms for encryption require two keys: a public and a private key, which form a key
pair. Algorithms are designed in such a way that anyone can encrypt a message using the public key, while only the
legitimate recipient, in possession of the private key, can decrypt the message. Because of the asymmetry between the two users of a public-key cryptosystem (one holding the private key, and keeping it secret, while the other user only needs to know a public, non-secret key, and worry about its authenticity), public-key cryptography is often referred to as asymmetric cryptography.

\paragraph{Secret key agreement based on public-key cryptography}
 As shown by Whitfield Diffie and Martin Hellman in 1976 \cite{DH76}, public-key cryptography can be used
to establish a shared secret key over an unprotected classical communication channel, without using a prior shared secret. It
thus provides a practical way to implement secret key agreement, in particular over open networks. Note however that Diffie-Hellman does not guarantee the authenticity of key agreement and thus that an additional authentication scheme needs to be provided. Over open networks such as the Internet, public-key infrastructures, i.e.\ trusted third parties, are often used for this purpose.

\paragraph{Security of public-key cryptography }
Public-key classical encryption schemes currently in use are based on well-studied mathematical problems that are believed to be computationally difficult to solve, such as the computation of discrete logarithms over finite fields or elliptic curves or the factorization of integers in the case of the RSA scheme.

In the case of RSA, it is necessary to use private and public keys of at least $1024$ bits, in order to offer a
reasonable security margin against the computational efforts of an eavesdropper and asymmetric keys of $2048$ bits are
preferable \cite{EcryptRec, Ecrypt08}. However, since the computational hardness of the underlying problems in public-key cryptography have not been formally established, public-key cryptography is not immune to scenarios where an eavesdropper would possess some unexpectedly strong computational power or would know better cryptanalysis techniques than
the best published ones. Moreover, most of the currently used public-key cryptographic schemes (for
example RSA) could be cracked in polynomial time with a quantum computer: this results from Shor's algorithm for discrete log and factoring, that has a complexity of $O(n^3)$ \cite{Shor}. It however seems possible to build alternative public-key cryptographic schemes on problems that could resist polynomial cryptanalysis on a quantum computer, such as lattice shortest vector problem \cite{Regev, Regev2}. Such schemes are nevertheless much less practical than RSA-like schemes. This topic is at the moment actively studied, in the framework of what is called post-quantum computing cryptography \cite{PQC09}, and we will discuss some implications of what researchers already know in subsection \ref{PQC}.

\paragraph{Speed of public-key cryptography}

Making the computations relative to the asymmetric cryptographic protocols requiring long keys, such as RSA, is a rather
computational intensive and time-consuming task. The performance of RSA-based key agreement implementations depends
heavily on hardware : for RSA 2048 implemented on an Intel Pentium IV with a 2.93 GHz processor, the computations needed for one key exchange (essentially one RSA encryption and one decryption) take roughly 10 ms \cite{CryptoPlus}. The same key exchange would be approximately ten times faster (thus in the ms range) on dedicated coprocessors and is ten times slower (in the time range of a few tens of a second) on smart card coprocessors \cite{RSAtime}. There are other public-key encryption protocols, for example based on elliptic curves, for which keys are significantly shorter (typically between 160 and 256 bits), but only slightly better speed performance can be obtained. As a consequence public-key cryptography is  too slow to be used in order to encrypt communications over data networks. Public-key cryptography is most commonly used solely for initial secret session key agreement (in  network protocols like SSL for example), while faster classical symmetric-key cryptography is then generally used for symmetric encryption and/or authentication of data.

 \subsection{Classical computationally secure symmetric-key cryptography and secret key agreement}  \label{subsec:sym}

Symmetric-key cryptography refers to cryptography methods in which both the sender and receiver share the same secret key. Symmetric-key encryption was the only kind of encryption publicly known until the discovery of public-key cryptography in 1976 \cite{DH76}.

Symmetric-key ciphers are used to guarantee the secrecy of the
encrypted messages. Modern study of symmetric-key ciphers
relates mainly to the study of block ciphers and stream ciphers
and to their applications. AES is a block cipher that had been
designed by a team of Belgium cryptographers (Joan Daemen et
Vincent Rijmen) and has been adopted as an encryption standard by
the US government (in replacement of DES). Block ciphers can be
used to compute Message Authentication Codes (MACs) and can thus
also be used to guarantee integrity and authenticity of messages.
Stream ciphers, in contrast to block ciphers, create an
arbitrarily long stream of key material, which is combined with
the plaintext bit-by-bit or character-by-character, somewhat like
the one-time-pad. We will however restrict our discussion to block ciphers in the 
remaining part of this sub-section, reference \cite{NESSIED20} provides a very complete survey of
classical computationally secure symmetric-key schemes.

\paragraph{Secret key agreement based on classical computationally secure symmetric-key cryptography }

Secret key agreement can be realized by using  solely
symmetric-key cryptographic primitives. For example, the combination of
a symmetric-key encryption scheme with a symmetric-key
authentication scheme  allows one to build a secret key agreement
primitive. Provided that an initial small secret key is previously shared,
symmetrically, by Alice and Bob, they can use a symmetric
cipher to encrypt messages. These messages (that can consist of random bit strings or not) will constitute the next keys that can be shared securely between Alice and Bob. The initially shared symmetric key material can be used to symmetrically compute (on Alice's side) and check (on Bob's side) a message authentication tag, and thus guarantee the authenticity of the newly distributed secret keys. As we shall see, $\mathrm{O(log}\, n)$ bits of secret keys are sufficient to authenticate $n$ bits of messages in this context. It is thus only necessary to pre-share a small initial quantity of secret keys, used for authentication to perform secret key agreement with symmetric primitives, therefore, only small initial secret keys are needed. One has to call such secret key agreement schemes {\it secret key expansion} schemes more than {\it secret key establishment} schemes.

\paragraph{Security of classical computationally secure symmetric-key-based secret key agreement.}

The security of secret key agreement based on classical symmetric-key
cryptography depends on the security of the cryptographic primitives that are used, and on the composability of those cryptographic
primitives. Shannon has proven that there is no unconditionally
secure encryption scheme which requires less encryption key bits than the one-time-pad \cite{Shannon}. This has a fundamental implication: the entropy of the encryption key needs to be at least as large as the entropy of the message to be encrypted if one wants to build an unconditionally secure scheme. Hence, if we consider the
possibility of building an unconditionally secure symmetric key
expansion scheme, i.e.\, a method to symmetrically generate secret
keys out of a short initial symmetric shared secret key, the former
results from Shannon imply that such a scheme is impossible to
achieve in the framework of classical  cryptography. However, as we shall see in subsection \ref{subsec:QKD}, such a cryptographic primitive is possible in a quantum cryptographic context.

It is however possible to use classical symmetric-key encryption
and authentication schemes, that are not unconditionally secure,
to build a secret key agreement scheme. AES can for example be used
for symmetric-key encryption and can be also used to compute
message authentication codes (using, for instance, CBC-MAC). The
security model that applies to such symmetric-key classical
encryption schemes (symmetric-key block ciphers and stream
ciphers) is not unconditional security (the entropy of the key is
smaller than the entropy of the message) and not even ``provably
computationally security''. The security model
that applies to classical symmetric-key cryptography can be called
``practical computational security'': a cryptographic scheme is
considered ``practically computationally secure'' if the
best-known attacks require too much resource (such as computation
power, time, memory) by an acceptable margin \cite{NESSIED20, Ecrypt08}.

There are no publicly known quantum attacks on classical
symmetric-key cryptographic schemes and the cryptanalysis of symmetric-key
classical cryptography on a quantum computer reduces to exhaustive
search. Here a quantum computer would thus still give an
advantage: performing exhaustive key search given a  known plaintext-ciphertext pair corresponds to the problem of finding a element  in a unsorted
database of $N$ elements. The complexity of this problem is of  $O(N)$ on a classical computer but
only of $O(\sqrt N )$ on a quantum computer \cite{Grover}. The complexity reduction offered by Grover algorithm is only polynomial (as opposed to the super-polynomial complexity reduction offered by Shor algorithm), and this implies that doubling the key size would be enough to maintain (against quantum computers) the level of algorithmic complexity one currently has today against classical computers for  symmetric-key primitives.

\paragraph{Performances}
Symmetric-key classical cryptography is by several orders of magnitude
 less computational intensive and thus leads to faster implementations compared to asymmetric
cryptography \cite{NESSIED21}.
There are now 128-bit AES encryptors able to
encrypt data at rates in the Gbit/s range \cite{Verbauwende}.  This is the reason why it is widely preferred to
use symmetric-key schemes for encryption and/or authentication
over currently deployed communication networks. AES is currently
the chosen standard for symmetric-key classical block ciphers.

Under the  assumption that there exists no better way to break a symmetric-key
cryptographic scheme is exhaustive search within the key
space (assumption that will be discussed in more details in subsection \ref{subsec:KeyAgeing}) 
then, a symmetric key of $103$ bits is roughly comparable,
in terms of computational requirements, to a RSA key
modulus of $2048$ bits. Note that doubling
the length of a symmetric key implies squaring the computational
efforts needed for exhaustive search; on the other hand, the
computational efforts do not scale as fast with key length in the
case of asymmetric cryptography \cite{Ecrypt08}.

\subsection{Quantum key agreement - quantum key distribution (QKD)} \label{subsec:QKD}
 
 \paragraph{Unconditionally secure key agreement relying on quantum physics}

 Quantum key distribution, invented in 1984 by Charles Bennett and
Gilles Brassard \cite{BB84} based on some earlier ideas of
Stephen Wiesner \cite{Wiesner} is a quantum cryptographic alternative solution to the secret key
agreement problem, between two users that trust each other, in the presence of an adversary. In contrast to public-key cryptography, it has been proven to be unconditionally secure, i.e.\, secure irrespectively of the computing power that may be used by an attacker \cite{MayersProof, BBBMR, ShorPreskillProof}.

An important consequence of the unconditional security of QKD is that it would remain secure even in the advent of a quantum
computer. On the other hand, legitimate users can perform unconditionally secure QKD even without possessing
themselves a quantum computer. QKD can thus be deployed today in order to secure communication networks. 

Rigorously speaking, quantum key distribution should be called quantum key agreement, or quantum key establishment, since the secret key shared at the end of the protocol is not decided upon solely by one of the player and then distributed to the other. However, as the expression ``quantum key distribution'' and the acronym QKD are now firmly established, we have chosen to stick to them throughout this article.

The existing work on (classical) secret key agreement by public discussion, studied in the
framework of information-theoretic cryptography \cite{Maurer93}, has played an important role in QKD security proofs and it is interesting to see that QKD has conversely also triggered some important development in classical information-theoretic cryptography \cite{BBCM95}. However, classical models are in general not sufficient to capture all the information that can be learnt by a quantum eavesdropper, and QKD security proofs must rely on elements of quantum information theory.

It is indeed possible to relate QKD security to the fact that it is impossible to gain information
about non-orthogonal quantum states without perturbing  these
states \cite{Peres, EHPP, BBM92}. This property is used to upper bound the amount of information that can have been gained by an eavesdropper, commonly called Eve, tampering on the quantum channel  connecting the two legitimate users, commonly called Alice and Bob. If the information bound on the eavesdropper is low enough, a key can be distilled between Alice and Bob, and this with perfect secrecy: the information Eve may have about the key is, with an exponentially high probability, below a vanishingly small upper bound.

In addition to the first general proofs mentioned above, the theory of QKD security has continued to evolve, with recent proof techniques that now refer to a security criteria that is composably secure  \cite{RennerPhD} and based on information measures (smooth min- and max-entropies,  that are generalization of the Shannon entropy) that give an operational framework to analyze the security of quantum (as well as classical) information-theoretic cryptography. This framework also allows to carry out the security analysis with a finite number of signals \cite{ScaraniRenner08}, as it was also the case in the earlier works of Mayers \cite{MayersProof} and Biham et. al. \cite{BBBMR}.

 \paragraph{Basic principles of QKD}

Without going into the details of the different implementations or
protocols (one can refer to Refs \cite{GisinRevModPhys,
LutkQC, ScaraniRMP} for an extensive overview on that matter) we can describe the structure and the principle of
operation of the basic practical QKD system: a QKD link.

\begin{figure}[!h]
\begin{center}
\includegraphics[width= 12cm]{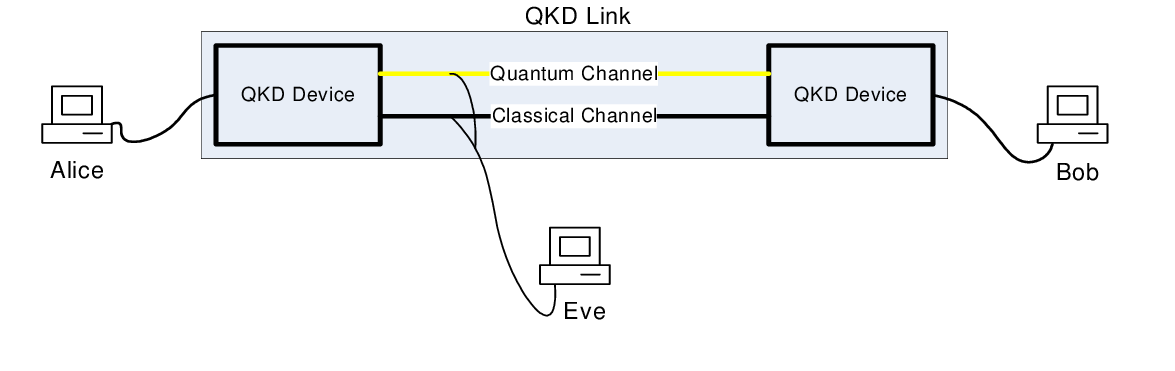}
\caption{Structure of a QKD link as it is referred throughout this
article} \label{fig:QKDLink}
\end{center}
\end{figure}

As depicted on Fig. \ref{fig:QKDLink}, a QKD link is a point-to-point
connection between two users, Alice and Bob, who
want to share secret keys. The QKD link is constituted by the
combination of a quantum channel and a classical channel.
Alice generates a random stream of classical bits and encodes them into
a sequence of non-orthogonal quantum states of light, sent over
the quantum channel. Upon reception of those quantum states, Bob
performs appropriate measurements leading him to share some
classical data correlated with Alice's bit stream. The classical
channel is then used to test these correlations. If the
correlations are high enough, this statistically implies that no
significant eavesdropping has taken place on the quantum channel
and thus that with very high probability, a perfectly secure
symmetric key can be distilled from the correlated data shared by
Alice and Bob. In the opposite case, the key generation process
has to be aborted and started again. This means in particular that any strong enough perturbation on the quantum channel, inducing noise above the security threshold (this threshold is for example of 11$\%$ for the BB84 protocol with one-way classical communications \cite{BB84, ScaraniRMP}), will in practice disrupt the key generation. An active attacker accessing the quantum channel can thus in practice mount denial-of-service (DoS) attacks and we will discuss in \ref{subsubsection: DoS} how such attacks can be mitigated in a network context.

 \paragraph{Authentication of the classical channel and everlasting security}

QKD is a symmetric secret key agreement technique that requires, as initial resources, a public quantum channel and an authenticated public classical channel. If one wants to stay in the strongest security model  i.e.\ the ITS paradigm, for the authentication of the communications on the classical channel, message authentication codes based on universal hashing can be used. Such authentication codes were first introduced by Wegman and Carter and further developed, especially by Stinson \cite{WC79, WC81, StinsonAuth}. In order to use ITS authentication in QKD, Alice and Bob need to share, in advance, a short secret key (whose length can scale almost logarithmically with the length of the secret key generated by a QKD session, for a given security parameter \cite{StinsonAuth}). QKD, operated in this regime, is an ITS symmetric secret key expansion scheme that has no classical counterpart since secrecy entropy cannot be increased by classical means \cite{MQR09}. Moreover, after initialization of the system (initial distribution of small secret authentication keys), authentication is not a burden for the global performance of QKD systems.

As discussed in \cite{VABook, Stebila09} and also in sections \ref{sec:MP} and \ref{sec:networks}, there are different ways to obtain an authenticated public channel, and non-ITS authentication primitives such as public-key authentication can be combined with QKD. In this case, the resulting QKD protocol is not strictly speaking unconditionally secure, but still verifies a very powerful security property called {\it everlasting security}: although the QKD-based key establishment now relies on some computational assumptions,  this potential weakness (against an adversary computationally very powerful) can only be exploited {\it during} the QKD protocol execution. If  the authentication mechanism is broken at any later point in time, it will not alter the security of the generated key. One can thus use QKD to build long-term unconditional secrets out of short-term secrets, hence the term everlasting security.

 \paragraph{Security assumptions in QKD}
 
 \label{par:secassump}
 Even though ``unconditional security'', synonym of information-theoretically, is a proper term to characterize the security of QKD, one must be careful with the precise meaning of this expression. As noted in \cite{GisinRevModPhys}, there are indeed several important underlying assumptions that must be fulfilled for QKD security proofs to be valid and the term ``unconditional'' can in a sense be misleading. Let us list three important assumptions:
 \begin{itemize}
 
 \item{[Quantum Mechanics] The security of QKD is intrinsically based on the assumption that Quantum Mechanics is correct. One important property is for example that non-orthogonal quantum states, onto which information is encoded in QKD, cannot be distinguished perfectly. This leads to a fundamental trade-off between the information that an eavesdropper can learn and the disturbance that it creates, which is exploited in security proofs.}
  
 \item{ [Secure labs] There  is no leakage of information from the honest parties labs of Alice and Bob. As a matter of fact, all the information about the secret key is processed in the labs of Alice and Bob and no key secrecy is achievable if this information leaks. This assumption may not be easy to enforce in practice. There are however technical and organizational measures (commonly used in security) that can help to back up this assumption: Alice and Bob QKD hardware can be put in tamper-resistant cases and / or stored in secure locations, with access control measures making it unaccessible to Eve.}
 
 \item{ [Trustworthy implementation] The implementation of Alice and Bob QKD devices is conform to what they expect it to be and to the modeling made in the security proof. This ``double trust assumption'' is directly challenged by the demonstration of successful attacks \cite{NatureMakarov11, LeuchsCalibration} on QKD implementations, as it will be discussed in \ref{subsec:sidechannels}. As a consequence, backing up this trust assumption on QKD implementation will imply to develop and implement counter-measures to known attacks and certification procedures for QKD devices in the same spirit as what exists today for (classical) cryptographic hardware. Another way to deal with this security assumption on QKD implementation might indeed be to go around it. This is the idea behind the so-called ``device-independent QKD protocols'' for which unconditional security can be proven independently of the hardware implementation of Alice and Bob QKD devices. }

The three generic security assumptions mentioned above are sufficient to set up the framework in which QKD security can be formally proven. As we shall see in \ref{subsec:sidechannels}, the last two assumptions can be challenged in the context of side-channel attacks, while on the other hand, the ``Trustworthy implementation'' assumption can be relaxed in the context of device-independent QKD. This last assumption, related to the trust in QKD hardware is fairly broad and must be made in principle for all the pieces of equipment inside a QKD device: laser source, phase/amplitude modulator, detectors, electronics, etc. There are nonetheless equipments that have a special status from a cryptographic point of view, namely random number generators (RNGs). RNGs are needed in most QKD implementations (this is not the case for QKD systems based on sources of entangled photon pairs) as local sources of presumably perfect entropy, i.e.\ totally unpredictable for an external observer. Most RNGs used today in computers and industrial products are pseudo-RNGs (PRNGs) that generate sequences of numbers that approximate the properties of random numbers but can be determined based on the knowledge of a small number of parameters. The use of PRNGs as entropy source is thus not possible in QKD, for it would contradict its information-theoretic security claims. On the other hand, there exist RNGs called true random number generators (TRNGs) where the entropy generated is based on physical processes. Randomness generation can for example rely on thermal noise in electric circuits (resistor, ring oscillator) is very hard to predict unless strong bias can be imposed on the noise process.  One might also design ``quantum random number generators'' (QRNGs) that are special TRNGs where the randomness is due to the unpredictable nature of the outcome of quantum measurements. Several quantum processes can be exploited to build a QRNG (nuclear decay, shot noise, reflexions of single photons on a semi-transparent mirror, etc.) and several QRNGs are now commercialized \cite{JenneweinQRNG, GisinQRNG} In order to perform QKD in a information-theoretic setting, TRNGs or QRNGs  should be used and incorporated in QKD implementations.  
   
 \end{itemize}

\subsection{Trusted couriers key distribution (TCKD)}

The trusted courier method is known since the ancient times: a
trusted courier travels between the different legitimate users to
distribute secret keys, hopefully without being intercepted or corrupted on his way by any potential opponent. Only practical
security can be invoked in this case, which has to be backed by
the enforcement of an appropriate set of security measures.
Although trusted couriers become costly and unpractical when
implemented on large systems, this technique has remained in use
in some highly-sensitive environments such as government
intelligence, or defense. 

The trusted couriers key distribution (TCKD)  is probably one of the methods used in the framework of network security for which the analogy with QKD is the closest:
\begin{itemize}
\item Like QKD, TCKD is a method relying on the physical security of the communication line between Alice and Bob, it is thus also sensitive to distance and other characteristics (danger, perturbations ...) of the communication line between Alice and Bob.
\item Like QKD, TCKD can be used as a secret key agreement protocol.
\item Like QKD, TCKD needs some initial trust in the relative identities of Alice and Bob. Moreover, like for QKD, this necessary initial trusted authentication can be handled via different techniques, such as the pre-distribution of a secret key (such as a password), or the use of an unforgeable (or at least reputed to be such) public identity certificate issued by a trusted third party (such as as the seal that was used by emperors and kings or the signed certificates we now use for public keys).
\item Like QKD, TCKD is a technique that currently finds its application when classical secret key agreement schemes are believed not to offer enough security guarantees.
\end{itemize}

Despite the similarities listed above, there are important differences between QKD and TCKD:
\begin{itemize}
\item The first difference is really intrinsic to QKD and TCKD
``physical realities''. In the case of QKD, the ``couriers'' are
quantum states of lights (flying qubits) traveling at the speed
of light and on which eavesdropping can be detected with arbitrary
high statistical certainty. On the other hand, TCKD cannot offer
any of those guarantees and, whether one uses human beings or
pigeons, trust or corruption of a classical courier cannot be
proven nor tested. 
\item Reliability, automation and cost
effectiveness will, very likely, be one of the major advances
offered by the development of QKD networks, that can moreover efficiently handle key management issues. On the other hand, reliability and cost of TCKD infrastructures are critical problems and there is no real hope that such systems can ever be automated, leading to serious key management issues and very high operational costs.

\item Unlike point-to-point QKD links, classical trusted couriers
are not intrinsically limited in distance. They are also not very
limited in rate since they can take advantage of the possibilities
offered by today's portable and versatile classical memories, such
as DVDs or USB keys, that can store Gigabytes of data. We will
however see in section \ref{sec:networks} that QKD networks could
be used to go beyond QKD links distance limitations and that such
networks could also be used to distribute secret keys ``on demand'' to the
end users, which is fundamentally different from relying on keys stored on the very same DVD, that could be duplicated at any later point in time if some adversary manages to break the protections around the storage device.
\end{itemize}

\subsection{Cascaded schemes and dual key agreement}

\paragraph{Cascaded ciphers}

The idea of {\it cascaded cipher} is to compose several encryption
primitives by applying them sequentially on the same cleartext.
Note that the encryption primitives can be of different types as
in AES-Twofish or the same one as in 3DES. The interest of
cascading ciphers is  to increase cryptanalysis' difficulty. As pointed out by Maurer and Massey, \cite{Massey},
the first encryption layer, i.e.\ the one directly applied to the
message, is in all cases the most important one.

\paragraph{Dual secret key agreement}
This idea of cascaded cipher can straightforwardly be applied to secret key agreement: two keys of the same length are established through two secret key agreement schemes (relying on either the same primitive or on different ones) and the final key is obtained by XORing these two keys. We will talk, in this context, of {\it dual secret key agreement}. Note that more than two secret key agreement schemes, of various types, can in principle be combined this way.  We will restrict in the following to a discussion of dual secret key agreement involving QKD as one of the secret key agreement technique.

The approach of dual secret key agreement could for example be beneficial when combining keys established through one classical key agreement scheme (RSA for example) and keys established through QKD: breaking the entire secret key agreement scheme implies breaking the classical key agreement scheme {\it and} breaking QKD. If one has doubts about the security of QKD, the dual secret key agreement procedure guarantees that the security will at least not be worse than that of the classical secret key agreement technique with which it is combined. The same is true if one has doubts about the security of secret key agreement scheme based on classical cryptography.
However, while there already exist security standards in classical cryptography (for example FIPS 140 \cite{FIPS} or Common Criteria \cite{CommonCriteria}), there are not yet such standards for QKD. The approach of dual secret key agreement could thus allow to certify a system according to already established criteria, without requiring to specify the quantum part of the key establishment. On the other hand, as we shall see in subsection \ref{QCert}, the certification of quantum cryptosystems is a topic on which work is already being initiated and we can hope to have FIPS-140 or Common Criteria certified QKD systems within a few years.

\section{Securing a point-to-point classical communication link by combining QKD with symmetric encryption} \label{sec:MP}

QKD is a secret key agreement primitive that can be performed at the physical layer level. In previous section, we have compared QKD to other existing solutions for secret key agreement. We will now analyze how the secret keys established by QKD can be used to perform a link layer cryptographic task: securing the data sent on a classical communication link, by relying on keys generated by QKD (plus some initially shared small secret authentication keys) and on symmetric-key cryptographic primitives.

More formally, we consider here the problem of securely transmitting  classical messages (payload) from Alice to Bob via the following generic protocol: 
\begin{enumerate}
\item Establishment of a symmetric secret key $K_S =  K_{encrypt} \cdot K_{auth}$ between Alice and Bob  ($X \cdot Y$ stands for the concatenation of string $X$ with string $Y$).
\item Secure and authentic transmission of the  message $M$ over the classical channel, with symmetric-key cryptographic primitives: $M$ is encrypted with encryption key $K_{encrypt}$ and authenticated with the authentication key  $K_{auth}$.
\end{enumerate}

After a brief subsection about the performances of QKD devices we will analyze several variations of the generic scenario described above, in which QKD is used as the secret key agreement primitive over a point-to-point link, while different types of encryption and authentication schemes are used.

\subsection{Performance of QKD link devices: recent progresses}

QKD research and development  is carried out on an international level \cite{ChinaQKD, SpaceQ, NISTGigabit2007, CanaryQKD, Sasaki2008, peev:inprep09} and QKD systems are being developed with increasing performances and reliability.   One can currently expect to exchange up to 1 Mbits of
secret key per second, over a point-to-point QKD link of 20 km \cite{dixon:oe08}. The maximum span of QKD links is now
roughly around 100 km or even 140 km \cite{dixon:oe08, stucki:qp08}  (depending on the type of single photon detector that is used) at 1550 nm on a telecom dark fiber. A comparable maximum span has also been reached in the context of  ground-to-ground free space QKD \cite{FS144}. This experiment was successfully realized with a quantum channel whose losses  were one order of magnitude larger than what we expect them to be in the framework of space-to-ground communications. It thus paves the way towards QKD between a satellite and a ground station \cite{SpaceQ}. Both secret bit rate and maximum reachable distance
are expected to continue their progression during the next years
due to combined theoretical and experimental advances. Note that
in any case QKD performances are intrinsically upper bounded by
the performance of classical optical communications. It is however also important to notice
that QKD systems can now basically be built with optimized,
off-the-shelves telecom components (laser, phase modulators,
beamsplitters, polarization controllers, and etc.) at the notable
exception of photodetectors. Photodetection is currently the bottleneck for the performance of QKD systems, but it is important
to keep in mind that, even on that side, although there are many technical problems to overcome, there are very few fundamental
limitations for rate and distance, as detection methods are making significant progresses \cite{LongDQKD, Cova2GHz, NISTGigabit2007, dixon:oe08, stucki:qp08, Sasaki2008}. Another approach, known as ``continuous variables QKD'' (CVQKD) uses only standard PIN photodiodes, but requires more sophisticated data post-processing in order to extract the secret keys \cite{QCV}. Significant progresses, on the theoretical \cite{RennerCirac2008} as well on the implementation side \cite{fossier:qp08}  have been achieved for CVQKD and further advances on the protocol side may allow CVQKD systems, that were known to be able to deliver high bit-rate but only for small or medium losses on the quantum channel, to become suitable for long-distance, high-bit rate QKD \cite{leverrier:qp08}.

As optical modulation and detection techniques become faster and faster (currently reaching GHz clock rates for discrete variable systems \cite{Cova2GHz}), data post-processing, and in particular error correction, may become the bottleneck in terms of reachable key rate.  As a consequence, serious efforts are invested to design fast and efficient implementations of error correcting codes adapted to the specificities of QKD. An initial protocol was introduced by Brassard and Salvail \cite{Cascade} in 1993: Cascade, and remains widely used.  Practical error correction technique like Cascade (whose decoding complexity is in $n \, log(n)$) or LDPC decoding (also in $n \,log (n)$) are both capacity achieving and efficiently decodable. LDPC are however more and more used in practical QKD, both for continuous \cite{JKL11} and discrete variable QKD \cite{MMEM13},  because of performance concerns (Cascade is interactive and can require a lot of network bandwidth) and also because the requirement of one-way classical communication (and therefore unidirectional reconciliation) is often a necessary hypothesis in existing security proofs.

\subsection{QKD composed with one-time-pad: long-term security of link encryption } \label{subsec:UClink}

One-time-pad encryption is the only encryption scheme for which information-theoretic security can be proven. It is thus natural to combine it with QKD.
Beyond the fact that it is ITS, QKD has the important property of being universally composable (UC) \cite{CanettiUC}. Universal composability of QKD can be proven \cite{RennerPhD} 
and implies that QKD can be composed with other ITS and UC protocols, resulting in a composed protocol that is also ITS and UC.
As a consequence, when keys established by QKD are used to perform one-time-pad encryption the resulting protocol is an unconditionally secure message transmission protocol \cite{ CompoOTP}.
Building an unconditionally secure classical communication link is probably one of the most important application of QKD. 

Taking advantage of the perfect secrecy offered by one-time-pad and from the fact that the keys established by QKD are
unconditionally secure, message encryption can be performed with a level of security that cannot be reached when the key agreement mechanism is not QKD: the messages are perfectly
secret with respect to adversaries and there is provably
absolutely no chance that future events  could alter the secrecy
of these messages. 


As pointed out in \cite{Stebila09}, long-term security is needed in many specific application scenarios, such as the protection of medical records, industrial secrets and military or governmental classified informations. 
However, offering long-term security for highly sensitive data is not something that can be guaranteed by today's computationally secure schemes. Indeed, as written in \cite{Ecrypt08}, ``beyond approximately 10 years into the future, the general feeling among ECRYPT partners is that recommendations made today should be assigned a rather small confidence level, perhaps in particular for asymmetric primitives''. 
As a matter of fact, it is important to note that when one deals with the transmission of encrypted information, an adversary can always store the ciphertext and wait for the decryption until better cryptanalysis methods become available (for example more efficient algorithms for factoring or the discovery of an efficient way to attack AES) or better cryptanalysis hardware (indeed large quantum computers would be very efficient for breaking most of the asymmetric encryption primitives in use today).
The recommendation of ECRYPT is indeed to consider using one-time-pad encryption for high-security levels, ``provided the key management can be solved'' \cite{Ecrypt08}. In this perspective, the combination of QKD with one-time-pad, which provides a practical solution for unconditionnally-secure data transmission over a point-to-point link (solution that can indeed be extended in the context of networks, see section \ref{sec:networks}) seems to be a natural response to meet some of the most stringent requirements within high-security communication infrastructures: long-term security.
 
\subsection{QKD composed with a classical computationally secure symmetric encryption scheme: key security and key ageing} \label{subsec:KeyAgeing}

Here we will consider one very frequent use case: QKD combined with link encryption performed with a symmetric encryption scheme (such as AES). This combination is the one that is currently pushed forward by existing commercial QKD vendors \cite{qkdstartups}. It provides a practical solution adopted within the BBN Darpa Quantum Network project \cite{BBN}. Such a composition provides a practical solution to realise a point-to-point VPN encryptor, that can be deployed in layer 2 (link) in the OSI network layer model \cite{qkdstartups} or directly in the layer 3 (network), for example by interfacing QKD-based key exchange with IPSEC \cite{BBN2, Dali}.

The final security of the exchanged data over such link cannot be stronger than the security of the encryption scheme. In the case of a symmetric-key block cipher, the security of the encrypted data depends on at least four factors:
\begin{enumerate}
\item the security of the encryption key (can an opponent get even some partial information about the key?);
\item the number of blocks that have been encrypted with the same key (key renewal rate);
\item  the length of the encryption key (56 bits for DES, 128, 192 or 256 bits for AES);
\item the security of the symmetric-key encryption algorithm, for which only ``practical computational security'' can be claimed.
\end{enumerate}

The last two factors are only dependent on the encryption technique and not at all on the key agreement scheme. The security implications (and the security level) associated with the choice of a given symmetric cipher, with a given key modulus length is discussed in detail in \cite{NESSIED20, EcryptRec, Ecrypt08}. In the ECRYPT Yearly Report on Algorithms and Keysizes published in july 2008 \cite{Ecrypt08}, a symmetric key modulus of 128 bits is recommended for long-term security (while 256 bits is recommended for a good protection of symmetric ciphers against a quantum computer).

The first two factors, on the other hand, are influenced by the choice of the secret key agreement scheme: the security of the key is intrinsically linked to the security of the secret key agreement scheme while the key renewal rate also strongly depends, on a practical level (hardware performance, security policy, implementation details, etc.), on the key establishment scheme. We will discuss in the following whether QKD-based schemes, used in replacement of traditional key agreement schemes, present an interest with respect to these two factors.

\subsubsection{Security of the key}
As explained in section \ref{sec:KD}, there exist many different solutions to perform secret key agreement, but QKD is the only existing and practically implementable scheme that can offer information-theoretic security.
One advantage of using QKD as the key renewal mechanism for link encryption is the long-term security guarantee for the keys. This is to be compared with the usual mode of operation for VPN encryptors, where the establishment of an encryption key relies on asymmetric cryptography and thus imply a vulnerability to potentially existing (now or in the future) computational attacks on the public-key scheme.

Another important operational interest of QKD, when used sequentially to produce successive encryption keys, is the property called {\it forward-secrecy} of the established keys:  the successive keys established over a QKD link are independent from one another. Therefore the potential compromise of a single key cannot lead to the compromise of other keys. We can  notice that the forward-secrecy of QKD is a natural consequence (and weaker property) than the everlasting secrecy mentioned in \ref{subsec:QKD}. As a matter of fact, in the sequential production of QKD keys, the secret material needed at each QKD round to authenticate the classical channel stems from a previous QKD round. 
Forward-secrecy in key establishment is an important property and can also be obtained with public-key cryptography under computational assumptions \cite{DHForwardSecure} while it cannot be obtained at all with computational symmetric cryptography since the successive keys are not independent from one another.

\subsubsection{Key renewal rate} \label{subsubsection:keyrenewalrate}

The rate at which encryption keys are renewed can influence the security of the encrypted data. This is what we call the {\it key ageing} factor, that can be reformulated as  a question:  how often secret session keys should be changed and what is the impact on the global security of the classical message passing scheme?
To give elements of answer to this question, we will consider the practical example of a link encryptor that corresponds to what current QKD vendors are selling: combining QKD-based secret key agreement with AES.

\begin{itemize}
\item A practical example: key renewal for AES encryption
\end{itemize}

Let's take the case of 128-bit AES for which Xilinx produces dedicated cipher modules that can support a data rate of $2.2$
Gbit/s and for which ``dedicated research hardware'' has demonstrated a rate of $21.54$ Gbit/s
\cite{Verbauwende}. In this case, the number of blocks (of 128 bits) encrypted per second (with a 128-bit key) is  $10^{8.23}
\simeq 2^{27}$ blocks/s. An exhaustive key search attack would in this case take $2^{101}$ seconds, i.e.\ roughly $8 \, 10^{22}$ years, way beyond the age of the universe ($\simeq 13 \, 10^{9}$ years), which means that exhaustive search attacks are not a threat to AES.

We must however not forget that the previous calculation is done under the assumption that exhaustive search is the best attack on AES.
It seems thus important to question this assumption and study what can be said about the influence of the encryption key renewal rate on the security of AES.
This complex question is intrinsically linked to the security assumptions one can make on AES itself.

\begin{itemize}
\item Security of AES
\end{itemize}

The cryptanalysis of encryption schemes like AES is a difficult topic that is still subject to very active research and it seems realistic to think that the ultimate difficulty of such cryptanalysis is currently not known. 

For block ciphers, the resistance to cryptanalysis depends in particular on the number of rounds applied when encrypting one block (see \cite{NESSIED20} for details).
Even though AES is considered secure and is currently a standard (in the USA for example, AES-128 is considered sufficient up to the SECRET security level, while AES-192 or AES-256 can be used for TOP SECRET communications  \cite{AESNSA}) it had been shown that weaker versions of AES, with reduced numbers of rounds, could be attacked successfully by strategies that require less computational efforts than exhaustive search \cite{NESSIED20}. Some cryptographers also claimed that powerful algebraic attacks could break AES based only on a very small number of known cleartext / ciphertext \cite{Courtois}. However, algebraic attacks have never been successfully demonstrated on AES and are not regarded as a real threat by the majority of the cryptography community \cite{Ecrypt08}.

Very recently in 2011, the first cryptographic break on the full version of AES has been published \cite{BogdanovAES}. This attack allows key recovery from AES-128 in $2^{126.1}$. Hence this result does not make mathematical attacks on AES really more practical than exhaustive search. It is however an incentive to renew AES encryption key relatively often, as we shall discuss in the next paragraph.

\begin{itemize}
\item Key renewal rate and use of QKD 
\end{itemize}

If AES is considered perfectly secure, then the limit of $2^{{\rm keylength}}$ blocks after which the keys have to be renewed in order to avoid collision-related problems is in practice not a problem, and one cannot justify the need to renew the AES keys several times per second as QKD can allow.

However, as we have seen in the previous paragraphs, there exist arguments, based on some known existing algorithmic weaknesses of reduced versions of AES, that indicate that it could be beneficial for the global security of AES encryption to refresh the secret keys after a number of blocks that is significantly smaller than $2^{{\rm keylength}}$. Moreover, as discussed in \ref{subsec:sidechannels}, in  the context of embedded systems where trust in the environment around the encryptor cannot be guaranteed, frequent key renewal can be beneficial (in addition to extra algorithmic and hardware security layers) because there exist efficient attacks that can break AES after a limited number of rounds. 

Finally, the most important justification to renew keys relatively often stems from the fact that, beyond the algorithmic security of the encryption algorithm, key management and key storage can be the main vulnerability in the security chain. In this perspective, renewing the keys often is a way to reduce the negative impact of a key leakage.

\section{Key agreement over a network of QKD links : QKD Networks} \label{sec:networks}

There are several fundamental limits regarding what can be
achieved with standalone QKD links. QKD links can by definition
only operate over point-to-point connections between two users,
which greatly restricts the domain of applicability of quantum key
distribution. Furthermore, QKD links  are limited in rate and distance, and cannot be deployed over any arbitrary
network topology. To overcome these limitations, it is
important to study what can be achieved by networking QKD links in
order to extend the extremely high security standard offered by
QKD to the context of long distance communications between
multiple users. The development of QKD network architectures
appears from this perspective as a necessary step towards the
effective integration of QKD into secure data networks.

We will begin this section by an overview on the different generic QKD network architectures that have already been proposed.
We will then present some elements of comparison between QKD networks and classical network, for the purpose of network-wide key agreement.

\subsection{QKD network architectures}

\label{qkdnetarch}

What we call  a ``quantum network''  is an
infrastructure composed of quantum links connecting multiple distant
nodes.  A quantum network can be used for key agreement, relying for that on QKD. We call  such infrastructures ``QKD networks''.

The essential functionality of the QKD network is to
distribute unconditionally secure symmetric secret keys to any
pair of legitimate users accessing the network. These first
elements of definition are however fairly generic and can be
refined. Indeed, even though we are at the infancy of the
development of QKD networks, different models of QKD networks have
already been proposed. 

It is convenient to characterise the different QKD network models
by the functionality that is implemented within the nodes and thus by the different underlying quantum network models. We can,
from this perspective, differentiate  three main categories of
network concepts, based on different ``families'' of node
functionalities:  1) optical switching;  2)  quantum relaying; and  3) classical trusted
relaying.

\paragraph{Optically switched quantum networks} These are networks in which  some
optical function, like beam splitting, switching, multiplexing,
demultiplexing, etc., can be applied at the network nodes on the
{\it quantum} signals sent over the quantum channel. The purpose
of such optical networking capabilities in the context of QKD
networks is that they allow to go beyond two-users QKD.
One-to-many connectivity between QKD devices was demonstrated over
a passively switched optical network, using the random splitting
of single photons upon beam splitters \cite{Barnett}. Active
optical switching can also be used to allow the selective
connection of any two QKD nodes with a direct quantum channel. The
BBN Darpa quantum network \cite{BBN, BBN2} contains an active
2-by-2 optical switch in one node, that can be used to actively
switch between two network topologies. Optical functions can thus
be used to realise multi-user QKD and the corresponding nodes do
not need to be trusted, since quantum signals are transmitted over
a quantum channel with no interruption from one end-user QKD
device to the other one. This QKD network model can however not be
used to extend the distance over which keys can be distributed.
Indeed, the extra amount of optical losses introduced in the nodes
will in reality shorten the maximum span of quantum channels.

\paragraph{``Full'' quantum networks} To extend the distance on which quantum key distribution can be
performed, it is necessary to fight against propagation losses
that affect the ``quality'' of the quantum signals as they travel along the quantum channel. Quantum repeaters\cite{Cirac} can
overcome the loss problem and can be used to form an effective
perfect quantum channel \cite{Biham}.  A quantum network where
nodes are constituted by quantum repeaters can thus be called a
``full'' quantum network. It is not necessary to trust the network
nodes to have unconditional security when performing QKD over such
full quantum networks.

 Quantum repeaters however rely on
elaborated quantum operations and on quantum memories that cannot
be realised with current technologies. As discussed in
\cite{Collins}, quantum nodes called quantum relays could also be
used to extend the reach of QKD. Quantum relays are simpler to
implement than quantum repeaters since they don't require
quantum memories. Building quantum relays remains however
technologically difficult and would not allow to extend QKD reach
to arbitrary long distances.

\begin{figure}[!h]
\begin{center}
\includegraphics[width= 14cm]{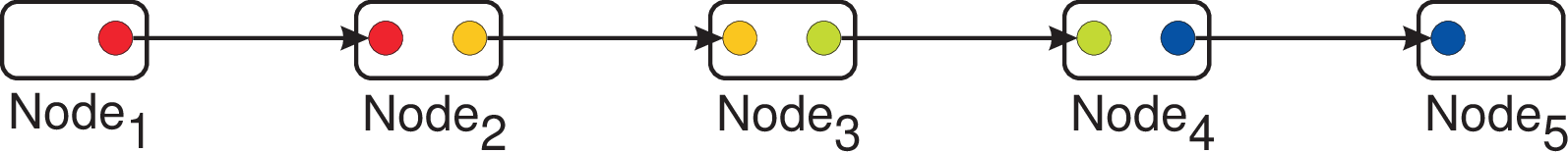}
\caption{``Hop-by-hop'' unconditionally secure message passing on
a path made of trusted relay nodes connected by QKD links. Message
decryption / re-encryption is done at each intermediate node, by
using one-time-pad between the local key, distributed by QKD,
$K_{local}$, and the secret message $M$ resulting in the ciphered
message $M \bigoplus K_{local}$. Different key associations are
symbolised by different colours.} \label{hop}
\end{center}
\end{figure}

\paragraph{Trusted repeater QKD networks} This technique where trusted classical memories are placed within the nodes can be implemented with
today's technology.  Trusted repeater QKD networks operation follow a
simple principle: local keys are generated over QKD links and then
stored in nodes that are placed on both ends of  each link. Global
key agreement is  performed over a QKD path, i.e.\ a
one-dimensional chain of trusted repeaters connected by QKD links,
establishing a connection between two end nodes, as shown on
Fig. \ref{hop}. Secret keys are forwarded, in a hop-by-hop
fashion, along QKD paths. To ensure secrecy, one-time-pad
encryption and unconditionally secure authentication, both
realised with local QKD key, are performed. End-to-end
information-theoretic security is thus obtained between the end
nodes provided the intermediate nodes can be trusted.

Classical trusted repeaters can be used to build a long-distance QKD network and real-scale deployments of such QKD networks have already been demonstrated, firstly with the DARPA Quantum network, deployed in Boston in 2004 \cite{BBN, BBN2} and more recently by the {\sc SECOQC} consortium that developed a highly integrated network architecture, with dedicated protocols, allowing global key management \cite{dianati:scn08, dianati:lcn07} and leading to a demonstration in 2008 in Vienna \cite{peev:inprep09}. A trusted repeater network is essentially a classical network where the exchanged data consists in keys encrypted with QKD-based keys. Since it relies on QKD for local secret key agreement and on unconditionally secure encryption, it can offer an unprecedented overall security for key agreement and can accommodate long-distances. This last claim is of course only true if one can guarantee that nodes can be trusted. Such assumptions is demanding but can be verified some existing high security networks infrastructures.  It can moreover be partially relaxed by making use of path diversity within networks, allowing to maintain security even when a limited number of nodes are corrupted \cite{salvail:jcs09, SandersRelay08}.

\subsection{ Key agreement in a network: elements of comparison between classical key agreement schemes and QKD networks} 

\subsubsection{Key Establishment Rate}

As discussed in \ref{subsubsection:keyrenewalrate}, some security requirements related to current block
ciphers such as AES could motivate the need to refresh secret keys of such ciphers over times shorter than a minute. Although
this is possible in practice with current technology, relying on Diffie-Hellman and PKIs, such key renewal rate policies are very seldom (if ever) enforced and
the key renewal period of most currently deployed VPNs is more in the range of hours. As a matter of fact, since public-key cryptography is rather slow and computational intensive and is using long key modulus  (see details in \ref{subsec:asym}), it could become an extremely high burden for end-users in terms of time and CPU consumption if key renewal was to be done over times shorter than one minute.
On the other hand, despite the fact that QKD is very often portrayed as slow \cite{WhyQC}, QKD rates, as we have mentioned earlier, are currently reaching several hundreds kbit/s for metropolitan distances. This implies that QKD networks could typically allow to refresh thousands of 128-bits AES keys per second, over VPN links in a metropolitan network.
\subsubsection{Network initialization and key pre-distribution}

Secure networks always rely on some initial trust in order to be able to provide some security guarantee. As discussed in  \ref{subsec:QKD}, a pair of initial small secrets or an authenticated classical channel is necessary to initialize a QKD link. We now consider the question of network initialization and key pre-distribution for symmetric-key-based and asymmetric-key-based secure networks and compare it with the requirements of QKD networks. As noted in \cite{Stebila09}, we can argue that the combination of asymmetric-key (for key pre-distribution) and QKD presents some specific interests.

\paragraph{Key pre-distribution over networks relying on symmetric-key cryptography}
One of the central issues in network key distribution is the initialization and the management of a potentially very large pool of secret keys: in a symmetric-key framework, where each member of an $n$-user network wants to be able to communicate securely with each of the other $n-1$ users, the key distribution scheme is required to provide any of the $n (n-1)/2$ pairs of users with a secret key before communication can start. Managing the security of those keys efficiently is thus a difficult task as $n$ grows. This is the reason why large-scale symmetric-key cryptography is seldom used in today's networks (however some network security schemes, like the Kerberos network authentication scheme \cite{Kerberos} rely on classical symmetric-key cryptography and on a single trusted center).

\paragraph{Key pre-distribution over QKD networks}

As pointed out in \cite{WhyQC}, QKD networks need pre-distributed
secret keys to perform the first rounds of authentication. The
QKD-generated keys can then be stored and used for later
authentication. Initialization of a QKD network of $n$ nodes thus
a priori requires the pre-distribution of $n (n-1)/2$ pairs of
secret keys (one per pair of user) with trusted couriers. One can however take advantage of network connectivity to show that it is sufficient to distribute keys over a subset of those
$n(n-1)/2$ pairs: what is needed is to distribute a pair of keys
over QKD links so that the resulting graph of ``initialized'' QKD
links is a covering graph of the QKD network. In this case, the
complexity of key pre-distribution only scales linearly with the network
size.

\paragraph{PKI Initialization}

PKI is the most commonly employed system for key agreement over
open networks. PKI trust relations are materialized by
certificates, i.e.\ signatures of public-keys and these trust
relations can be organized hierarchically, which offers the
advantage that trust relations do not have to extend over all nodes in the network but only to a trusted third party called the
certification authority. Public-key cryptography allows to
perform secure key agreement between two users without any pre-shared common secret: the only condition is
that they accept to trust the same certification authority. PKIs however also need to be
initialized, which might require the use of trusted couriers. In this sense, the initialization
of a QKD network and the initialization of a PKI are two problems that share some similarities.

\paragraph{Interest of PKIs for QKD Network Initialization}

As pointed out in \cite{WhyQC}, QKD networks present a
security advantage over PKIs when we consider the initialization
phase: in order to threaten the security of a QKD network, message
authentication needs to be broken {\it before or during} the execution of the quantum key establishment protocol. This property is connected to the everlasting security of QKD explained in \ref{subsec:QKD}.

One can take advantage of this
property in the case of QKD network initialization and consider an
hybrid scenario for key pre-distribution in which the classical
communications needed for the  key distillation phase are
authenticated, at least during the first QKD sessions, by a
computationally secure message authentication scheme based on
public-key cryptography (for which the PKI has been freshly
initialized). If no active attack on authentication has been
performed {\it before} the first potentially vulnerable (assuming some potential weaknesses of asymmetric cryptography) QKD
sessions, then the keys shared by Alice and Bob are identical and
unconditionally secure.  The previous condition will
always be verified if the computational power of the adversary is
bounded {\it at the time of the QKD network initialization} which constitutes a moderate assumption.
There is a clear practical interest for such a scheme: it relaxes the requirement of distributing pre-established small keys in a
QKD network for each network initialization (which requires secret couriers and can be a difficult key management problem in the case
of large networks).

\subsubsection{Vulnerability against denial-of-service attacks}
\label{subsubsection: DoS}

As discussed in \ref{subsec:QKD}, an individual QKD link is vulnerable to denial-of-service (DoS) attacks that can be brutally realized by physically cutting the quantum channel, or, more subtly, by acting on the quantum channel in order to increase the noise above the security threshold, so that no key can be generated despite the availability of the QKD link.

As we are interested here in comparing classical or quantum networks, it is important to note that the vulnerability to DoS attacks targeting individual communication lines, exists in both cases. There is an important difference: physical signals in classical networks can be broadcast, while QKD optical signals sent on the quantum channel can't, as this would violate the no-cloning theorem. Broadcasting the same quantum signals over multiple channels can thus not be used in QKD networks to resist against DoS attacks on some channels. 

On the other hand, XORing keys obtained via key agreements realized over disjoint paths (connecting the same endpoints) is a solution that can be used both in a classical and in a quantum context and that can be employed to perform reliable classical communication against an opponent performing DoS attacks on a limited number of paths. In this case, the resilience against DoS attacks, characterized by the maximum number of coordinated attacks that can be tolerated (while still being able to perform key agreement), depends on the topology of the network. 

There is another important aspect to consider: since the classical communications in a QKD protocol require the use of keys for authentication, DoS attacks can be designed to exhaust the key reservoirs on Alice and Bob and thus make the QKD link unusable to generate more key, unless it is reinitialized with small authentication keys. In a network context, there are however several solutions to key exhaustion: path diversity can be used, as explained above, and if there exists one functional path in the QKD network connecting the two endpoints of the link, this path can then be used to rekey the link. Moreover, besides the obvious but costly solution of using trusted courier to rekeying exhausted key store, one could also use the hybrid solution mentioned above, i.e.\ using a PKI for the first authentication round.

\subsection{Open networks versus trusted QKD networks}
\label{subsec:openvstrusted}

As pointed out in \cite{EcryptChallenges}, ``quantum cryptology is
not a solution for open networks'', i.e.\ a QKD network does not
allow users that do not share any pre-established secrets or trust
relations  to exchange a key and then communicate securely. In a
sense QKD networks are also tied by their ``physical nature'': they are limited in distance since quantum signals are 
exchanged over lossy optical channels. These physical limitations
however bring a considerable security advantage: QKD networks can
provide unconditional security to all the users that have access
rights to the network and are thus inside the ``circle of trust''
of these closed networks. As we shall discuss in \ref{PQC}, the practicality of asymmetric cryptography and its suitability for open networks may have as a counterpart the existence of stronger vulnerabilities to cryptanalysis. 


The difference between quantum networks and classical networks
thus appears to be almost ontological : they do not offer the
same services and exhibit a relation with space and distance that
is extremely different. While classical open networks, and
especially the Internet have been described as ``small worlds'',
where physical signals can be regenerated, data can be copied and
distances are almost abolished, quantum networks
are in essence closed networks where distance comes back into the
game, as it is the case with telephone networks \cite{alleaume:qp09}.

\section{Challenges and future directions } \label{sec:futureD}

As shown by the discussions we have conducted in sections 3 and 4, QKD is essentially adapted to be used in combination with classical cryptographic techniques. To move forward towards larger adoption, several factors will play an important role, in particular QKD system performances (in terms of key rate or reachable distance), their security, as well their integrability into modern optical networks. Since this paper focuses on the cryptographic status of QKD, we will essentially concentrate on challenges related to the practical security of QKD, and will evoke some research topics where the interplay between classical and quantum cryptography is likely to open interesting new perspectives.



\subsection{Practical security of QKD implementations and implementation loopholes}
\label{subsec:sidechannels}

\subsubsection{Physical side-channels}

Instead of trying to break the theoretical foundations of a given cryptographic system, another ``attack philosophy'' is to attack its implementation via loopholes, in order to gain some secret information via unconventional channels such as electromagnetic radiation, heat dissipation, acoustic noise, observation of computation time or power consumption.

As first demonstrated in the pioneering work of P. Kocher at the end of the 1990's, one can exploit so-called ``physical side-channels'' through which information about secret keys leaks while cryptographic computations are being conducted to mount extremely powerful attacks. The first efficient attacks to be demonstrated were based on monitoring the execution time and then the power consumption of classical cryptographic devices (whose implementations ultimately rely on semiconductor logic gates  and transistors) \cite{KocherTime, KocherDPA}. The variety of side-channel attacks and of counter-measures has gradually expanded since then and one can for example consult the ECRYPT Side Channel Cryptanalysis Lounge for a thorough overview \cite{SideChannelLounge}.

 The threats imposed by side-channel attacks can be in practice extremely serious. Their study and the development of countermeasures has thus become an increasingly important topic, in particular in the context of embedded cryptographic systems evolving in potentially hostile environments, such as smart cards. If we  consider for example the vulnerabilities of AES to side-channel attacks, state-of-the-art DPA attacks  (Differential Power Analysis attacks, i.e.\ passive attacks performed by monitoring the power consumption of the device) can successfully break unprotected AES implementations on a smart card after the acquisition of $100$ power traces, while roughly $50 000$ power traces are needed to break software protected implementations of AES. Even protected implementations of AES can thus be broken in a few minutes with DPA\cite{AESCHES08}. Moreover, if one allows active attacks, such as attacks based on fault injection in the circuit, then a full break of AES128 has been obtained in a few seconds, with only $2$ pairs of correct and faulty ciphertext \cite{DFA08}. Note also that implementation of asymmetric cryptographic schemes, such as RSA, are also vulnerable to side-channels attacks \cite{RSAPellegrini}. To increase security, dedicated counter-measure need to be deployed, as well as extra layers of hardware protections (in order to physically restrict the possibility to launch such attacks).
 
As we shall explain in more details in the next paragraph,  QKD is not immune to side-channels attacks and it has already been demonstrated that some specific implementation imperfections in QKD systems can also be exploited to mount side-channel attacks.

\subsubsection{ Quantum hacking}

As explained in \ref{subsec:QKD}, some trust assumptions are always needed to prove the security of a QKD protocol. In particular, Alice and Bob's hardware is assumed to be in secure labs (the [Secure labs] assumptions) and its operation is supposed to conform to what Alice and Bob expect ([Trustworthy Implementation]).
When one deals with real QKD implementations, both of these assumptions, and thus the entire security proof can however be challenged if some imperfections can be exploited to mount side-channel attacks. This line of research, called {\it quantum hacking} has become an important research field during the past years, where different types of attacks on QKD implementations have been proposed and experimentally demonstrated. After tackling the general question of the categorization of (passive or active) attacks on QKD systems, we will briefly review the main side-channel attacks studied so far and comment on the general situation of QKD with respect to practical security.

\paragraph{Passive and active side-channel attacks in QKD} By analogy with the attacks on classical cryptographic systems, it is interesting to distinguish between passive and active side-channel attacks in QKD. This categorization is not straightforward since, unlike in the classical case, ``passive listening''  on a quantum signal creates disturbances.
This leads to the question of whether the concept of passive side-channel attack is meaningful in QKD. However, since disturbances due to eavesdropping on a quantum signal are taken into account by security proofs, the corresponding attacks  should not be considered side-channel attacks. We can, on the other hand, define side-channel attacks as attacks exploiting discrepancies between the security proof framework and the actual characteristics of the implementation.

\begin{itemize}
\item Passive side-channel attacks correspond to attacks where the attacker does not actively modify the characteristics of the implementation, and tries to exploit existing imperfections to break the security. This can for example be possible if some information leaks from Alice's lab, as in the very first QKD experiment performed in 1989 \cite{BBBSS92}, where the Pockels cells used to modulate the polarization on Alice's side made noises indicating the state sent by Alice. This lead Charles Bennett to notice that the QKD system was only secure against an eavesdropper who happened to be deaf.
\item Active side-channel attacks correspond to attacks where the opponent actively modify some characteristics of the implementation in order create an exploitable deviation between the way the system under attack operates and the assumptions in the security proof. Most of the attacks evoked in the remaining part of this paragraph such as blinding attack and Trojan horse attacks are active side-channel attacks.
\end{itemize}

\paragraph{Review of the main side-channel attacks}

\begin{itemize}

\item {\it ``Conventional'' side-channels}. Most side-channels studied so far in the QKD research literature are ``quantum-specific loopholes''. This focus is natural since the main difference between QKD links and classical cryptosystems stems from the fact that the quantum channel is fully under the active control of Eve, leading to specific security issues.  However, ``conventional'' side-channels, i.e.\ security loopholes related to non-quantum aspects of the QKD protocol are also a serious threat. The accumulated knowledge from classical cryptography, regarding the secure design of cryptographic hardware will thus be extremely precious in this domain. One important class of conventional side-channels are the covert channels leaking information about the internal state of Alice or Bob's QKD hardware, for example via acoustic or electromagnetic radiations. The TEMPEST \cite{Tempest} measurement standards, dedicated to ensure protection requirement of cryptographic hardware against compromising emissions, are therefore meaningful in order to enforce the cryptographic boundary around Alice and Bob's QKD devices ([Secure lab] assumption).

\item {\it Trojan horse attacks}. Trojan horse attacks on QKD implementations \cite{Vakhitov2001} consist in sending Alice or Bob, via the quantum channel, intense pulses of light that will reflect back on the optical elements. By monitoring the reflected light, information about  the internal state of QKD devices can be learned by an eavesdropper. In particular, information about the status of the Alice's modulators (phase of polarization) can be learned and thus information about the key. Counter-measure against Trojan horse attacks can consist in putting spectral filters on Alice and Bob's side, as well as an optical isolator on Alice's side, and also to actively monitor the incoming light in Alice and Bob's devices.

\item {\it Attacks on single photon detectors}. Single photon detectors, usually avalanche photodiodes (APDs) operated in trigger mode are one of the key elements in discrete-variable QKD implementations. Many of the published attacks have been realized by exploiting existing or induced imperfections of APDs. In the {\it detector blinding attack} \cite{MakarovNJP09, NatureMakarov11}, an intense pulse of light is used to change the detector response characteristics, allowing then to use faked-states sent by Eve to mimic a correct behavior \cite{FakedStates, makharov:qic08} (with respect to the observation of correct detection statistics as well as correct correlations between Alice and Bob's measurements) while breaking the security of the QKD protocol since the detector response is manipulated by Eve. Many implementations rely on several APDs (for example one per detection basis in BB84) and a lack of symmetry between the detectors can be exploited to launch attacks. A temporal efficiency mismatch between two APDs can for example be used by the eavesdropper to launch the so-called {\it time-shift attack}: by adding a random temporal delay to the pulse, the eavesdropper can obtain information about  which detector has clicked and thus about the key \cite{MakarovPRA06, LoPRA08}. A countermeasure against the {\it time-shift attack} is to bring back the symmetry by performing a 4-state modulation (instead of 2-state) on Bob side \cite{Fung09}.

\item {\it Side-channel on random number generators}. If Eve is able to learn some information about the randomness generated by Alice and Bob to run the QKD protocol, this could lead to a complete security breach. As already discussed in \ref{subsec:QKD}, true number generators (TRNGs), whose entropy is intrinsically based on quantum processes should be used in QKD implementations, however a classical cryptographic processing of the output of the generator is useful to guarantee the robustness of the TRNG against some external physical biases. In addition, TEMPEST protection should also be applied to the QRNGs.

\item {\it Attacks on calibration procedures}. In a practical QKD system, some important parameters which play a role in the security of the system (for example the intensity of the laser source, the length of the quantum channel or the timing of the gate for single-photon detectors), are not directly measured during the ``quantum phase'' of the QKD protocol, but are calibrated. It means they are  measured offline: possibly once for all in the factory; or regularly if this parameter can fluctuate (this for example is the case for the gate delay in the Clavis2 implementation of BB84 by IdQuantique). The calibration protocols used to measure such parameters are thus important for security and it has recently been demonstrated that attacking the calibration of a QKD system can lead to security breaches \cite{LeuchsCalibration}. Calibration protocols in QKD must thus be considered in an adversarial setting. It is also true in the case of continuous-variable QKD, for parameters such as shot-noise measurements, that may be altered if the eavesdropper is left free to actively manipulate the local oscillator \cite{GrosshansCLEO}.

\item {\it Attacks on multi-photon pulses - Extended security models}. In the BB84 protocol and other discrete-variable protocols, many among the early security proofs \cite{MayersProof, BBBMR, ShorPreskillProof} assumed that Alice sends single-photon pulses.
However, single photon sources are experimentally challenging to make and thus impractical to use in real QKD systems. Instead, weak coherent laser pulses (WCP) with poissonian photon statistics are used. This has security implications since pulses of light containing more than one photon can be exploited by the eavesdropper to learn potentially all the information about the encoded bit, using the so-called {\it photon-number-splitting (PNS) attack} \cite{BLMS00}. Considering this attack, it remains possible to prove the security of BB84 performed with WCP, but one must decrease the mean number of photons in the pulse (and hence the rate) as the distance increases \cite{BLMS00}. In 2003, a more radical response to the PNS attack  has been proposed, that consists in modifying the QKD protocol with the adjunction of decoy-states to actively test the influence of the eavesdropper on the photon statistics. This protocols allows to obtain a linear decrease of key rate with distance, which is optimal \cite{Hwang, HKLo}. 

More generally, it has been shown that small deviations between the model and the implementation, such as imperfect preparation of the states by Alice or imperfect detection by Bob, can essentially be captured in the security model, at the expense of some extra privacy amplification \cite{GLLK}. Note that is not the case for attacks like blinding or Trojan horse, where the deviation from the security model is not small.

\end{itemize}

\subsection{Device-independent security: fundamental quantum mechanics as a tool against side-channels}

Device-independent security \cite{AcinDeviceInd} is a promising development of quantum cryptography, with no classical counterpart. It allows to perform cryptographic tasks, with unconditional security, without assuming that the underlying hardware implementation is trusted. Device-independent security is based on the non-locality of entangled quantum states. Moreover the non-local correlations involved in Bell Inequalities (BI) violation, reachable with an entangled quantum state \cite{GisinRevModPhys}, can be related to the {\it absence of side-channels}: one can test and verify that the Hilbert space in which the quantum state of the system is controlled and observed is not leaking information towards another Hilbert space and thus to a potential
eavesdropper \cite{Gisin2006}. As shown in Ref. \cite{AcinDeviceInd} it is possible to build  so-called device-independent QKD protocols that are intrinsically resistant to side-channel attacks. However, for device-independent security proofs to be valid, BI violations must be tested in the so-called loophole-free regime, which remains currently an experimental challenge. As a consequence, practical long-distance device-independent QKD seems out of reach for the moment (but is significantly easier than large scale quantum computing).

It is in essence impossible to prove such side-channel-resistance properties with classical cryptographic systems, because any
classical message can be duplicated and cloned without any perturbation. It is fascinating to notice that some
very deep aspects of quantum information tools,  like the loophole-free Bell Inequalities testing \cite{Grangier2001}, that
happen to be at the heart of quantum theory foundations, are seemingly bound to play an important role in the future
development of secure cryptographic hardware.

\subsection{Cryptographic certification of quantum cryptosystems}
 \label{QCert}
 
One aspect that may strongly hinder the adoption of QKD in network security has been spotted a long time ago by Michael Nielsen \cite{NielsenBlog}: quantum
cryptosystems are currently lacking one important element: {\it historical security}. The confidence in a cryptosystem indeed not only stems from the fundamental principles of its security, but also from the fact that it has been intensively tested, attacked and verified by a large number
of experts and users. It is clear that current QKD implementation cannot claim a large historical security, since few commercial systems are available and only few teams around the world have tried to attack the potential weaknesses of QKD systems. However, as explained in \ref{subsec:sidechannels} the work on quantum hacking as well as the development of counter-measure is gaining a lot of momentum, thus improving the situation.

In parallel to this necessary scientific research on practical security, the emergence of security standards and certification procedures for QKD device will be a necessary step to build a security referential against which systems can be tested. This work, which had started within the {\sc SECOQC} project \cite{DCCC03}, is now lead within a dedicated working group ETSI, the European Telecommunications Standards Institute. The QKD ISG, industry standardization group, launched in October 2008, brings together important actors from science and industry in order to converge towards industry standards for QKD, including standards concerning the cryptographic certification of QKD hardware implementations. 

\subsection{Post-quantum computing cryptography}
\label{PQC}

As noted in \cite{EcryptChallenges}, ``If powerful quantum computers could be built, most asymmetric cryptographic protocols in use today would no longer be secure, which would present a serious challenge for open networks''. As also noted in \cite{EcryptChallenges} and explained in \ref{subsec:openvstrusted}, QKD cannot constitute a solution for open networks either, however the potential impact of quantum computation on cryptography is motivating collaboration between the quantum information and the cryptography community, with an impact on the development of both fields.

Beyond the classical ITS key establishment schemes discussed in \ref{subsec:CITKE}, the fast-growing knowledge accumulated on quantum computation must be taken into account when designing new public-key schemes and study their resilience to quantum computing attacks. An important topic is thus to find computational problems that would be difficult on a quantum computer, and on which public-key schemes could be based. The lattice shortest-vector problem \cite{Regev} on which the NTRU public-key scheme is built or the McEliece public-key encryption system, based on coding theory \cite{McEliece}, both fall in this category of (likely) quantum-resistant public-key cryptosystems.

Such QC-resistant public-key schemes probably constitute the most plausible solutions to replace the current asymmetric schemes, if large quantum computers were made, or if problems such as factoring or discrete logarithm extraction were to be broken. However, public-key schemes like McEliece is significantly less efficient, in terms of key sizes,  than current public-key schemes while the security of NTRU is still not considered well established.
 It is also important to notice that even though there exists no known quantum algorithm able to attack efficiently  NTRU or McEliece public-key schemes, none of them are proven to be difficult problems (of super-polynomial complexity) on a quantum computer (indeed there is no proof that they are difficult on a classical computer).  On one hand, the lattice shortest-vector problem, which is NP-hard on a classical computer, seems to be of a comparable complexity on quantum computer \cite{Regev2, Mosca13} which is an indication that quantum computers could not solve this problem efficiently. On the other hand, the fact that NTRU is based on the shortest-vector problem (SVP) does not imply that the only way to break NTRU is to solve SVP.
 
Indeed, as noted in \cite{Simon}, the question of whether the complexity classes related to mathematical problems are ultimately different on a quantum computer than what they are classically remains essentially an open question. There are even some indications leading to partially negatively answer to this question: as demonstrated in \cite{BBBV}, oracle methods can be used to give evidence that the complexity class NP  is not included in BQP (which contains the problems that can be considered as efficiently solvable on a quantum computer), somehow setting a limit to the power of a quantum computer.
 
Understanding algorithmic security in a world where large-scale quantum computers may exist one day implies to take a fresh look on cryptographic designs and is connected to fundamental questions in computational complexity. Post-quantum cryptography is thus an extremely important and stimulating research field, not only for the cryptography of tomorrow, but also for the cryptography of today. 

\subsection{Classical Cryptographic Primitives built on top of QKD networks}

QKD networks operated with trusted repeaters can be deployed today and can therefore be considered as a new security infrastructure.  We believe that it could be fruitful to also consider such networks from a purely theoretical point of view, as ``new cryptographic primitives'', allowing the establishment of unconditionally secure keys, among a network of trusted centers connected by QKD links.

It seems interesting to examine what new classical
cryptographic protocols could be built on top of such networks,
beyond global pair-wise key establishment. As already proposed in \cite{DSEC17} in the framework of the
bounded quantum-storage model \cite{SalvailBounded}, QKD networks could be combined with Oblivious Transfer in order to allow unconditionally secure
multi-party computations. One can also study the efficiency of
secret sharing schemes over such new cryptographic infrastructure.

Important work has already been done on that topic (totally
independently from QKD networks considerations) \cite{DarcoRobust, Darco2006, DW02, DDWY93}. This work strikingly fits with the unconditional security offered by QKD networks, and powerful
information-theoretic tools have been developed to guarantee the
security of such networks even when some fraction of the network
nodes are corrupted. We believe that this opens promising research
perspectives in the domain of unconditionally secure networks.

\section{Conclusion}

QKD is currently the only known cryptographic technique that has lead to secret key agreement protocols for which the unconditional security can be formally established. Since the first QKD protocol, BB84, proposed 25 years ago \cite{BB84}, prolific theoretical and experimental research work has been conducted, as illustrated by our survey mostly centered on European realizations. Quantum cryptography has rapidly become an established academic topic within quantum information science, while QKD technologies have continuously moved forward in terms of performance and reliability.


The acknowledgement of these advances by security experts and by leading classical cryptographers is likely to play a key role in the development dynamics of a QKD industry and cannot be taken for granted. The main objective of this article was thus to give an overview of the development of QKD, as a cryptographic technology, with an emphasis on practical scenarios such as its use for key renewal in order to realize link encryption and its deployment at a network scale. The general message is that the most interesting uses arise when the long-term confidentiality of QKD-established keys can be exploited to provide some security advantage that could not be attained otherwise. This is in particular the case when QKD is combined with one-time-pad encryption, while it could also be the case when QKD is combined with AES. Concerning its deployment in a network context, the rationale for the use of QKD is to focus on medium-sized operated (closed) networks and also to carefully manage the established keys as well as the authentication procedure. One interesting feature is that, when combined with public-key infrastructures, used for initial authentication, the use of QKD networks then provides long-term security of the established keys.
 
Concerning future developments of QKD, we believe that it can benefit from cross-disciplinary approaches on fundamental topics such as the construction of ITS network protocols  or the study of  side-channels in cryptographic hardware. Synergies with network security research and with industry are likely to play an increasing role as certification procedures for QKD devices are adopted and products tested on real-world networks. We believe that the search of long-term security should be the driving concern when integrating QKD into security infrastructures. 

\section*{Acknowledgments}

We would like to thank collegially all the partners of the {\sc SECOQC} project for stimulating discussions and contributions to this article. R. A. thanks Gilles Van Assche for his help on improving subsection \ref{subsec:KeyAgeing}. R. A also acknowledges enlightning discussions with Sylvain Guilley, Philippe Hoogsvorst and Jean-Luc Danger about side-channel attacks on cryptographic hardware. Finally, R.A. warmly thanks Thomas Lawson, Damian Markham and Delphine Agut for their help on the manuscript.
We acknowledge support from the European Union under project {\sc SECOQC} (IST-2002-506813). R. A. acknowledge support from Agence Nationale de la Recherche under projects {\sc PROSPIQ} (ANR-06-NANO-041-05), {\sc SEQURE} (ANR-07-SESU-011-01) and {\sc COQC} (ANR-08-EMER-003).  R. A. also acknowledges support from the FP7 Marie Curie project Q-CERT 251467.

\end{document}